\documentclass[useAMS,usenatbib]{mnras}
\usepackage{aas_macros}
\usepackage{gensymb}
\usepackage{amsmath}
\usepackage{amssymb}
\usepackage{float}
\usepackage{graphicx}
\usepackage{lscape}
\usepackage{url}
\usepackage{natbib}
\usepackage{placeins}
\usepackage{color}
\usepackage{wasysym}
\usepackage{hyperref}
\usepackage{scrextend}
\usepackage{quotes}
\usepackage{afterpage}
\usepackage{placeins}
\usepackage{balance}

\newcommand{\nhii}{\ensuremath{n_{{\mbox{\small{H}\sc{ii}}}}}}
\newcommand{\nhi}{\ensuremath{n_{\mbox{\small{H}\sc{i}}}}}
\newcommand{\nH}{\ensuremath{n_{{\mbox{\small{H}}}}}}

\begin{document}

\title[Effects of Charge Exchange on HD~209458b]{Effects of Charge Exchange on the Evaporative Wind of HD~209458b}

\author[A. Debrecht et al.]{Alex Debrecht$^{1}$\thanks{adebrech@ur.rochester.edu}, Jonathan Carroll-Nellenback$^{1}$, Adam Frank$^{1}$,
\newauthor
Eric G.~Blackman$^{1}$, Luca Fossati$^{2}$, Ruth Murray-Clay$^{3}$, John McCann$^{4}$ \\
$^{1}$Department of Physics and Astronomy, University of Rochester, Rochester NY 14627\\
$^{2}$Space Research Institute, Austrian Academy of Sciences, Schmiedlstrasse 6, A-8042 Graz, Austria\\
$^{3}$Physics and Astronomy Department, University of California, Santa Cruz, Santa Cruz, CA 95064\\
$^{4}$Department of Physics, University of California, Santa Barbara, Santa Barbara, CA, 93106\\
}

\date{}

\pagerange{\pageref{firstpage}--\pageref{lastpage}}
\maketitle
\label{firstpage}

\begin{abstract}
The role of charge exchange in shaping exoplanet photoevaporation remains a topic of contention. Exchange of electrons between stellar wind protons from the exoplanet's host star and neutral hydrogen from the planet's wind has been proposed as a mechanism to create "energetic neutral atoms" (ENAs), which could explain the high absorption line velocities observed in systems where mass loss is occurring. In this paper we present results from 3D hydrodynamic simulations of the mass loss of a planet similar to HD~209458b. We self-consistently launch a planetary wind by calculating the ionization and heating resulting from incident high-energy radiation, inject a stellar wind into the simulation, and allow electron exchange between the stellar and planetary winds. We predict the potential production of ENAs by the wind-wind interaction analytically, then present the results of our simulations, which confirm the analytic limits. Within the limits of our hydrodynamic simulation, we find that charge exchange with the stellar wind properties examined here is unable to explain the absorption observed at high Doppler velocities.
\end{abstract}

\begin{keywords}
hydrodynamics -- planet-star interactions -- planets and satellites: atmospheres -- planets and satellites: individual: HD 209458b
\end{keywords}

\section{Introduction}

The amount of research being done on exoplanetary atmospheres continues to increase, especially with the pending launch of the James Webb Space Telescope. The detection of biosignatures and the analysis of atmospheric structure and composition have been prominent; however, the interactions of planetary atmospheres with the surrounding environment are also important. These interactions, especially atmospheric escape caused by the absorption of stellar X-ray and extreme ultraviolet radiation, play a key role in determining the long-term evolution of planetary atmospheres.

One of the most interesting phenomena involving evaporating planets is the absorption of Lyman-$\alpha$ radiation by neutral hydrogen out to large values of Doppler shift, approximately $\pm 150$ km/s. (For a somewhat more detailed discussion of studies on planet-star interactions, see \citet{debrecht20}.) The processes that have been suggested as causes for this high-velocity absorption include confinement by the stellar wind \citep{Schneiter2007, Schneiter2017, mccann18}, acceleration of neutrals by radiation pressure from stellar Lyman-$\alpha$ emission \citep{lecavelier08, Schneiter2017, bourrier13, khodachenko17, cherenkov18, debrecht20}, and charge exchange between the stellar and planetary winds \citep{holmstrom2008, ekenback10, tremblin13, kislyakova2014, bourrier16, christie16}. The differing conclusions resulting from these studies show that it is not yet clear which combination, if any, of these processes can fully account for the observations. Here we address this question by focusing on one process: charge exchange with the host star's wind.

It has been proposed that charge exchange between stellar wind protons and neutral hydrogen of planetary origins could create a population of high-temperature neutral hydrogen whose thermal and/or bulk velocities would be sufficient to absorb Lyman-$\alpha$ at Doppler shifts of the required high velocities. A number of studies have supported this idea. Most recently, \citet{khodachenko19} performed a low-resolution 3D parameter space study using a self-consistent hydrodynamic model of GJ~436b's wind-wind interactions and found a set of parameters that reproduced the observed absorption signature. \citet{bourrier16} performed a similar 3D study of GJ~436b using a particle model, finding that a significantly slower, denser wind than \citet{khodachenko19} provided the best fit to the observed absorption. Studies of HD~209458b in 2D with a wide range of stellar wind densities find absorption of only a few percent at high velocities, not quite enough to explain the observed level of absorption \citep{christie16,shaikhislamov16,khodachenko17}. An exception here is \citet{tremblin13}, who launched a wind from 4 $R_p$ (i.e. not self-consistently) in a high-resolution 2D simulation and found absorption of about 10\%, which is comparable to the observed absorption signal. In addition, \citet{esquivel19} have combined radiation pressure and charge exchange in a 3D simulation with an imposed planetary boundary condition at $3 R_p$ and no self-shielding for the Lyman-$\alpha$ radiation pressure and find that the combination may be sufficient to explain the observed absorption profiles.

In this paper we present results from high-resolution 3-D hydrodynamic simulations of a planet similar to HD~209458b, launching the wind self-consistently by calculating the ionizing radiation, injecting a stellar wind, and allowing the stellar and planetary winds to exchange electrons. In section \ref{sec:meth} we present the computational method and parameters used in our simulations. In section \ref{sec:result} we show the results of our simulations and perform an analytic estimate of the efficacy of charge exchange. In section \ref{sec:disc}, we compare our results to the analytic calculation and to the results of previous studies and suggest possible reasons for our lack of a high-velocity absorption signature. We conclude in section \ref{sec:conc}.

\section{Methods and Model} \label{sec:meth}

Our simulations were conducted with AstroBEAR\footnote{https://astrobear.pas.rochester.edu/} \citep{cunningham09,carroll13}, as described in \citet{debrecht20}, section 2. The ionization and recombination equations have been modified to account for separation between the stellar and planetary species:
\begin{equation}
 \frac{\partial {\nhi}_*}{\partial t} + \boldsymbol{\nabla} \cdot ({\nhi}_* \boldsymbol{v}) = (\mathcal{R} - \mathcal{I}) \frac{{\nhi}_*}{\nhi} + \mathcal{X},
\end{equation}
\begin{equation}
 \frac{\partial {\nhi}_p}{\partial t} + \boldsymbol{\nabla} \cdot ({\nhi}_p \boldsymbol{v}) = (\mathcal{R} - \mathcal{I})\frac{{\nhi}_p}{\nhi} - \mathcal{X},
\end{equation}
\begin{equation}
 \frac{\partial {\nhii}_*}{\partial t} + \boldsymbol{\nabla} \cdot ({\nhii}_* \boldsymbol{v}) = (\mathcal{I} - \mathcal{R}) \frac{{\nhii}_*}{\nhii} - \mathcal{X},
\end{equation}
\begin{equation}
 \frac{\partial {\nhii}_p}{\partial t} + \boldsymbol{\nabla} \cdot ({\nhii}_p \boldsymbol{v}) = (\mathcal{I} - \mathcal{R}) \frac{{\nhii}_p}{\nhii} + \mathcal{X},
\end{equation}
where ${\nhi}_*$ is the number density of neutral stellar ("hot") hydrogen, ${\nhi}_p$ is the number density of neutral planetary ("cold") hydrogen, $\nhi = {\nhi}_* + {\nhi}_p$ is the total number density of neutral hydrogen, ${\nhii}_*$ is the number density of stellar ionized hydrogen, ${\nhii}_p$ is the number density of planetary ionized hydrogen, $\nhii = {\nhii}_* + {\nhii}_p$ is the total number density of ionized hydrogen, $\mathcal{X}$ is the charge exchange rate, and $\mathcal{I}$ and $\mathcal{R}$ are the ionization and recombination rates.

Radiation pressure has not been included in these simulations ($F_{0,\alpha} = 0$) in order to isolate the effect of charge exchange on the Lyman-$\alpha$ absorption.

\subsection{Charge exchange} \label{sec:ch_exch}

Charge exchange is performed as in \citet{christie16}, where the charge exchange rate
\begin{equation} \label{eqn:chrg_exch}
    \mathcal{X} = \beta ({\nhii}_* {\nhi}_p - {\nhi}_* {\nhii}_p),
\end{equation}
and $\beta = 4\times10^{-8} \mbox{ cm}^3 \mbox{s}^{-1}$ is the charge exchange rate coefficient, which is the product of the charge exchange cross section and the relative velocities of the stellar and planetary hydrogen \citep{tremblin13}. This is equivalent to the charge exchange implementation in \citet{tremblin13}, and is electron-conserving.

\subsection{Planet atmosphere model} \label{sec:atmo}

We have modeled the planet as a sphere of hydrogen in hydrostatic equilibrium as in \citet{debrecht20}; refer to section 2.2 for details.

\subsection{Stellar wind}

Stellar winds are generally modeled isothermally. However, because our simulations use a polytropic index of $\frac53$, inserting a traditional isothermal stellar wind would result in nonphysical computational effects. In order to overcome the same problem, \citet{mccann18} derived a model for the stellar wind with a non-isothermal equation of state. We use the same model, with our stellar wind described by the following equations:
\begin{equation}
    \rho(r) = \frac{\dot M_\star}{4 \pi v(r) r^2},
\end{equation}
\begin{equation}
    T(r) = \frac{m_H P_0}{k_B \rho_0 (1+X)}\left ( \frac{v_0 r_0^2}{v(r) r^2}\right)^{(\gamma - 1)},
\end{equation}
\begin{equation}
    P(r) = \frac{k_B \rho(r) T(r) (1+X)}{m_H},
\end{equation}
\begin{equation}
    \phi(r) = - \frac{G M_p}{|a - r|} - \frac{G M_\star}{r} - \frac12 \Omega^2 r^2
\end{equation}
\begin{equation}
\begin{split}
    \frac12 (v(r)^2 - v_0^2) + \frac{\gamma}{(\gamma - 1)} \left(\frac{P(r)}{\rho(r)} - \frac{P_0}{\rho_0}\right)\\ + \phi(r) - \phi_0 = 0,
\end{split}
\end{equation}
where the final equation implicitly provides us with $v(r)$. Here $r_0$ is the reference radius, $r$ is the distance from the star, $X$ is the ionization fraction, and $\gamma$ is the polytropic index. Note that $P(r)$ can be written in terms of only $v(r)$. The stellar wind is specified by a combination of $r_0$, $v_0$, $T_0$, and $\dot M_\star$ (see Table \ref{tab:runs}). $\rho_0 = \rho(r_0)$, $P_0 = P(r_0)$, $\phi_0 = \phi(r_0)$. See \citet{mccann18}, Appendix A.3 for details of the derivation.

Our reference radius is well outside the simulation domain. We therefore provide a summary of relevant properties of the wind in the simulation domain here. For the low density wind, the number density $n_\star = 600 \mbox{ cm}^{-3}$, the radial velocity $v_\star = 70$ km/s, and the temperature $T = 1.4\times10^6$ K; for the high density wind, $n_\star = 4\times10^4 \mbox{ cm}^{-3}$, $v_\star = 45$ km/s, $T = 1.6\times10^6$ K; for the solar-analogue wind, $n_\star = 1.4\times10^4 \mbox{ cm}^{-3}$, $v_\star = 200$ km/s, $T = 7.7\times10^5$ K.

\subsection{Description of simulation}

We have run two simulations, varying the density of the stellar wind between each. See Table \ref{tab:runs} for stellar wind parameters, and section 2.3 of \citet{debrecht20} for details of the planetary parameters and simulation domain. The high density stellar wind was injected into the simulation domain from all but the $-y$ boundaries, with the most significant ingress from the $-x$ and $+y$ boundaries. The low density stellar wind had insufficient pressure to displace the ambient, so instead was introduced initially by replacing material below the initial ambient density with the appropriate stellar wind parameters. After introducing the stellar wind, we ran the low-wind simulation for 2.03 days and the high-wind simulation for 2.29 days, after which the simulations had reached a steady state, by which we mean the flow had achieved a stable ionization front and wind morphology.

In addition, a simulation with a solar-analogue wind was run in order to compare the effects of bulk and thermal velocities on synthetic observations. This simulation was run for 0.9 days. Though it appears to nearly reach a quasi-steady state, computational restrictions prevented this simulation from being run for multiple crossing times.

\setcounter{table}{0}

\begin{table*}
\centering

 \caption{Run parameters}
 \label{tab:runs}
 \begin{tabular}{l|c|c|c|c|c|}

  \hline

  & & No Wind & Low Density Wind & High Density Wind & Solar Analogue Wind \\

  \hline
  
  Planet Radius & $R_p\,(R_J)$ & \multicolumn{4}{c}{$1.529$} \\
  Planet Mass & $M_p\,(M_J)$ & \multicolumn{4}{c}{$0.73$} \\
  Planet Temperature & $T_p\,(\mbox{K})$ & \multicolumn{4}{c}{$3\times10^3$} \\
  Planet Surface Density & $\rho_p\,(\mbox{g cm}^{-3})$ & \multicolumn{4}{c}{$1.625\times10^{-15}$} \\
  Stellar Mass & $M_\star\,(M_{\astrosun})$ & \multicolumn{4}{c}{$1.23$} \\
  Stellar Radius & $R_\star\,(R_{\astrosun})$ & \multicolumn{4}{c}{$1.19$} \\
  Stellar Ionizing Flux & $F_{0,UV}\,(\mbox{phot cm}^{-2}\mbox{s}^{-1})$ & \multicolumn{4}{c}{$2\times10^{13}$} \\
  Stellar Wind Radius & $r_0\,$ (cm) & N/A & $4\times10^{11}$ & $4\times10^{11}$ & $5.5\times10^{11}$ \\
  Stellar Wind Velocity & $v_0\,$ (cm/s) & N/A & $2\times10^7$ & $2\times10^7$ & $2.3\times10^7$ \\
  Stellar Wind Temperature & $T_0\,$ (K) & N/A & $1.35\times10^6$ & $1.35\times10^6$ & $1\times10^6$  \\
  Stellar Wind Mass Loss Rate & $\dot M_\star\,(M_{\astrosun} \mbox{ yr}^{-1})$ & N/A & $2.6\times10^{-17}$ & $1.3\times10^{-15}$ & $2.378\times10^{-14}$ \\
  Orbital Separation & $a$ (AU) & \multicolumn{4}{c}{$0.047$} \\
  Orbital Period & $P$ (days) & \multicolumn{4}{c}{$3.525$} \\
  Orbital Velocity & $\Omega$ (rad/day) & \multicolumn{4}{c}{$1.78$} \\
  Polytropic Index & $\gamma$ & \multicolumn{4}{c}{$\frac53$} \\
  
  \hline

 \end{tabular}
\end{table*}

\subsection{Assumption of collisionality} \label{sec:knud}

\begin{figure*}
\centering
\includegraphics[width=\textwidth]{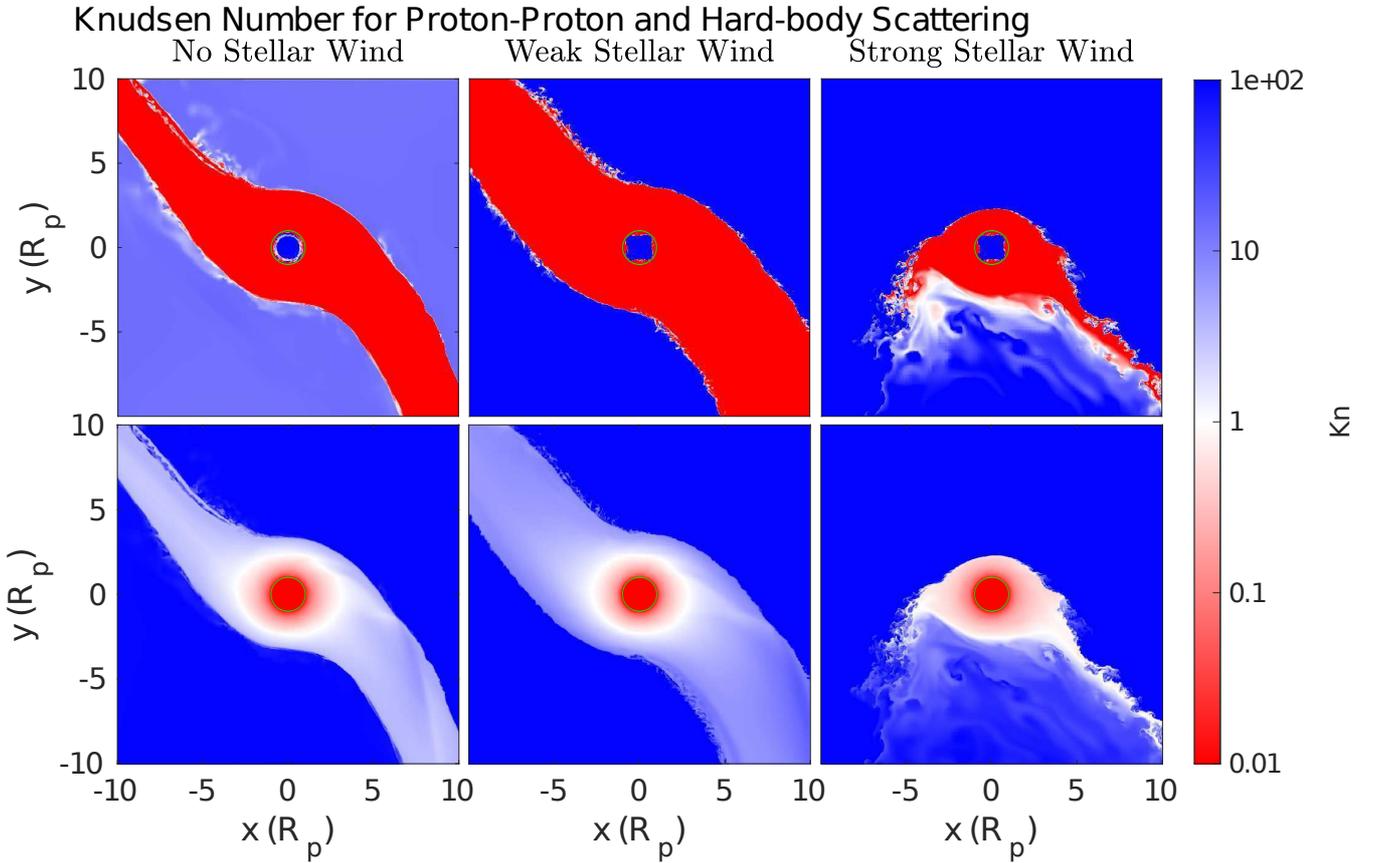}
\caption{Plots of the Knudsen number for the final states of our no-flux, intermediate-flux, and high-flux simulations. The top row shows the Knudsen number when the collisional cross-section is assumed to be due to Coulomb collisions, while the bottom row shows the Knudsen number when the cross-section is assumed to be due to hard-body collisions. The top row shows that the planetary wind is collisional due to ion-ion interactions (and therefore the stellar wind interactions with the planetary wind are collisional, though the stellar wind itself is not), while the bottom row shows that the extended planetary atmosphere is collisional due to hard-body collisions.}
\label{fig:knudsen}
\end{figure*}

We check the assumption that the fluid is collisional by plotting in the way described in \citet{debrecht20}, see Figure \ref{fig:knudsen}.

In each simulation, the planetary wind itself is highly collisional due to proton-proton interactions, while the planet's atmosphere is collisional due to neutral interactions. However, the stellar wind is significantly lower density than the planetary wind, and is therefore not collisional in general. Despite this, its interactions with the planetary wind are essentially collisional, thanks to the high density and ionization fraction of the planetary wind and the large cross section for charge exchange (see section \ref{sec:cross_sec}).

\section{Results} \label{sec:result}

\subsection{Analytic treatment} \label{sec:math}

An equilibrium state can be derived by determining at what number density of hot neutrals the charge exchange rate goes to zero. Our constraints are provided by the fact that the total number densities of planetary material, stellar material, neutral material, and ionized material are constant:
\begin{align}
    {\nhi}_p + {\nhii}_p = {\nH}_p = {\nhi}_{p,0} + {\nhii}_{p,0}\\
    {\nhi}_\star + {\nhii}_\star = {\nH}_\star = {\nhi}_{\star,0} + {\nhii}_{\star,0}\\
    {\nhi}_p + {\nhi}_\star = {\nhi} = {\nhi}_{p,0} + {\nhi}_{\star,0}\\
    {\nhii}_p + {\nhii}_\star = {\nhii} = {\nhii}_{p,0} + {\nhii}_{\star,0},
\end{align}
where the right-hand side represents the state before the interaction of the stellar and planetary winds (i.e. pre-equilibrium; in fact, the right-hand side could represent any pre-equilibrium state). Note that one of these equations is redundant, so that our system is still underspecified. Our final constraint is that in equilibrium, the charge exchange rate is 0, so that (see equation \ref{eqn:chrg_exch})
\begin{equation}
    {\nhii}_p {\nhi}_\star = {\nhi}_p {\nhii}_\star.
\end{equation}
This gives equations for the equilibrium state of each species:
\begin{align*}
    {\nhi}_p = \frac{{\nhi}_{p,0}^2 + {\nhi}_{p,0}{\nhi}_{\star,0} + {\nhi}_{p,0}{\nhii}_{p,0} + {\nhii}_{p,0}{\nhi}_{\star,0}}{\nH}\\
    {\nhii}_p = \frac{{\nhii}_{p,0}^2 + {\nhii}_{p,0}{\nhi}_{p,0} + {\nhi}_{p,0}{\nhii}_{\star,0} + {\nhii}_{\star,0}{\nhii}_{p,0}}{\nH}\\
    {\nhii}_\star = \frac{{\nhii}_{\star,0}^2 + {\nhii}_{\star,0}{\nhii}_{p,0} + {\nhi}_{\star,0}{\nhii}_{\star,0} + {\nhi}_{\star,0}{\nhii}_{p,0}}{\nH}\\
    {\nhi}_\star = \frac{{\nhi}_{\star,0}^2 + {\nhi}_{\star,0}{\nhi}_{p,0} + {\nhi}_{p,0}{\nhii}_{\star,0} + {\nhi}_{\star,0}{\nhii}_{\star,0}}{\nH}.
\end{align*}

Sampling the neutral tail in the no-wind case, at approximately (7.5, -2, 0), ${\nhi}_{p,0} = 3.1\times10^4 \mbox{ cm}^{-3}$, ${\nhii}_{p,0} = 5.4\times10^5 \mbox{ cm}^{-3}$. The weak stellar wind has ${\nhi}_{\star,0} = 7.2\times10^{-6} \mbox{ cm}^{-3}$ and ${\nhii}_{\star,0} = 7.1\times10^2 \mbox{ cm}^{-3}$. In equilibrium this gives us
\begin{align*}
    {\nhi}_p = 3.1\times10^4 \mbox{ cm}^{-3}\\
    {\nhii}_p = 5.4\times10^5 \mbox{ cm}^{-3}\\
    {\nhii}_\star = 6.7\times10^2 \mbox{ cm}^{-3}\\
    {\nhi}_\star = 3.8\times10^1 \mbox{ cm}^{-3},
\end{align*}
while the strong stellar wind has ${\nhi}_{\star,0} = 2.4\times10^{-2} \mbox{ cm}^{-3}$ and ${\nhii}_{\star,0} = 3.8\times10^4 \mbox{ cm}^{-3}$, which gives us
\begin{align*}
    {\nhi}_p = 2.9\times10^4 \mbox{ cm}^{-3}\\
    {\nhii}_p = 5.4\times10^5 \mbox{ cm}^{-3}\\
    {\nhii}_\star = 3.6\times10^4 \mbox{ cm}^{-3}\\
    {\nhi}_\star = 1.9\times10^3 \mbox{ cm}^{-3}.
\end{align*}

We can calculate an approximate optical depth as a function of frequency. Since the hot neutral density is much more significant in the high stellar wind case, we use it to compute the estimate. From Figure \ref{fig:charge_exchange_pot}, we see that the width of the interaction region between the stellar and planetary winds is approximately $1 R_p$. Since the hot neutrals are primarily directed down-orbit, we take the line-of-sight velocity to be zero. The optical depth, with thermal broadening, is then given by
\begin{equation}
    \tau(\nu) = \sigma_{\nu_0} {\nhi}_\star d \frac{c}{\nu_0} \sqrt{\frac{m_H}{2 \pi k_B T}} e^{-\frac{(\nu - \nu_0)^2 c^2 m_H}{2 \nu_0^2 k_B T}}.
\end{equation}
At $\pm 100$ km/s and $10^6$ K, this gives an optical depth of 0.061, or absorption of $\sim 1\%$.

We expect charge exchange equilibrium to be reached on the order of seconds:
\begin{equation}
    \tau_{CX} = \frac1{n \sigma_{CX} v_{th}} \approx 1,
\end{equation}
where the thermal velocity $v_{th}$ is approximately 150 km/s. Since these simulations happen over a number of days, we expect that charge exchange equilibrium will be maintained once we reach quasi-steady state.

\subsection{Simulation results}

All of our simulations are centered on the planet and carried out in the planet's orbiting frame of reference, with the orbital velocity vector $\boldsymbol\Omega$ in the $+z$ direction. Therefore, "up-orbit" is approximately in the $+y$ direction and "down-orbit" is approximately in the $-y$ direction. The no-wind case was discussed in \citet{debrecht20}; Figures \ref{fig:rxt_low_wind} and \ref{fig:flow_texture_low_wind} are exceedingly similar, being produced by a stellar wind calculated to insignificantly perturb the initial steady state. Here we summarize the low-wind, high-wind, and solar-analogue wind cases. Movies for each of these cases are available on the AstroBEAR YouTube channel\footnote{https://www.youtube.com/user/URAstroBEAR \label{ftn:movies}}.

\subsubsection{Low-density stellar wind}

\begin{figure*}
\centering
\includegraphics[width=\textwidth]{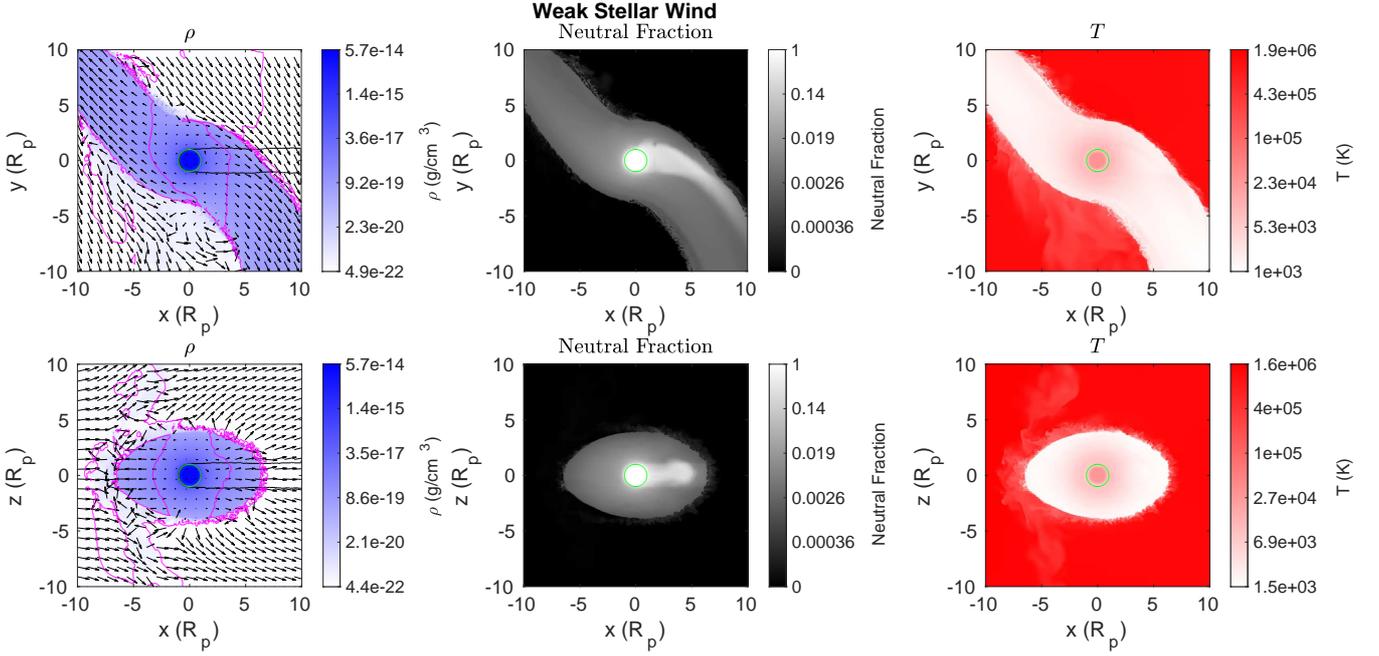}
\caption{The quasi-steady state of the low stellar wind case. The top row shows the view looking down on the orbital plane, while the bottom row shows the view standing in the orbital plane and looking up-orbit. The left column shows density, with the magenta contour the Mach surface and the black contour the $\tau = 1$ surface and the vectors giving the direction of the velocity. The green contour is the location of the nominal planetary radius $R_p$. The center column shows the neutral fraction, and the right column shows the temperature. The star is located to the left of the simulation grid. The planetary wind has expanded slightly thanks to the lower pressure of the stellar wind compared to the no-wind ambient. The stellar wind creates Kelvin-Helmholtz instabilities along the edges of the planetary wind.}
\label{fig:rxt_low_wind}
\end{figure*}

\begin{figure*}
\centering
\includegraphics[width=\textwidth]{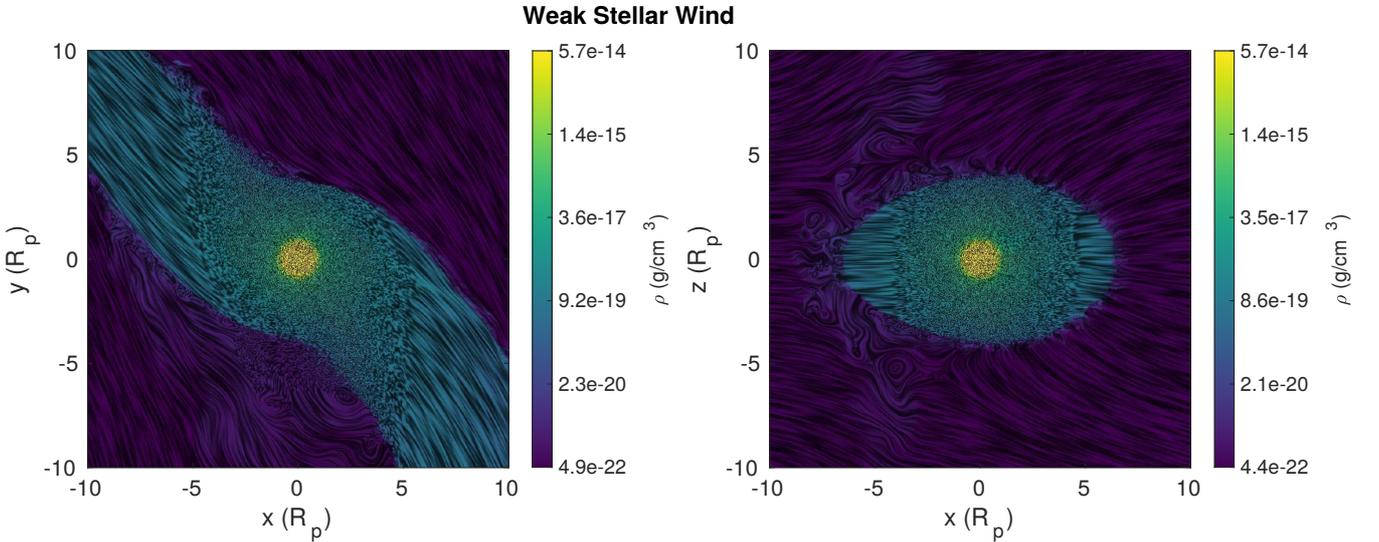}
\caption{Flow texture plot of the low stellar wind case, with the left panel showing the view looking down on the orbital plane and the right panel showing the view standing in the orbital plane looking up-orbit. The hue represents density, and the texture represents velocity streamlines. The star is located to the left of the simulation grid.}
\label{fig:flow_texture_low_wind}
\end{figure*}

Figures \ref{fig:rxt_low_wind} show the steady state of the low-wind run, which uses a stellar wind calculated to insignificantly perturb the initial steady state of the planetary wind. It can be seen that this case differs only slightly from the initial steady state, as intended. The most significant difference between the low-wind and no-wind states is the increased width of the torus of material, due to the lower total pressure of the stellar wind compared to the ambient medium used prior to its injection. The stellar wind interacts primarily along the edges of the wind, producing Kelvin-Helmholtz instabilities and, in the case of the down-orbit side of the planetary wind, drags material off into a low-density cloud. The neutral tail does not interact much with the stellar wind here. The stagnation region between the stellar and planetary wind below the down-orbit arm can be seen in Figure \ref{fig:flow_texture_low_wind}. For an in-depth discussion of the initial steady state, see section 3.2.1 of \citet{debrecht20}.

\begin{figure*}
\centering
\includegraphics[width=\textwidth]{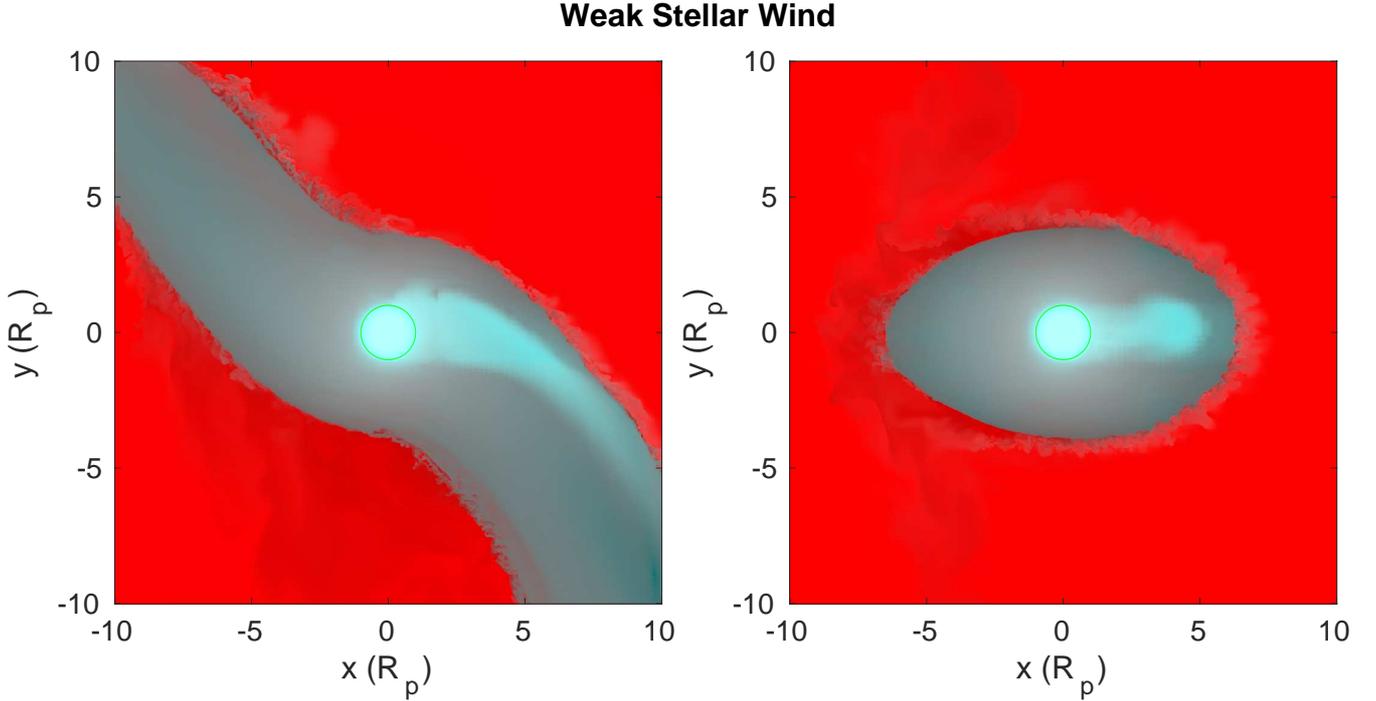}
\caption{An RGB figure where the red channel is temperature and the green and blue channels are the neutral fraction. Therefore, pure red represents stellar ions, white represents stellar neutrals, teal represents planetary neutrals, and black represents planetary ions. We can see that there are few stellar neutrals, which are concentrated in the instabilities along the edge of the wind. The neutral tail of planetary material is also clear.}
\label{fig:charge_exchange_low_wind}
\end{figure*}

Figure \ref{fig:charge_exchange_low_wind} displays the relative concentrations of the four species using a three-channel image. The red channel represents temperature. Since we only have two properties of interest, both the green and blue channels represent neutral fraction. Therefore, hot neutral material is white, hot ionized material is red, cold neutral material is teal, and cold ionized material is black. The hot neutral material of interest should be close to white. As expected from the calculations in section \ref{sec:ch_exch}, we find very little hot neutral material in this simulation.

\subsubsection{High-density stellar wind}

\begin{figure*}
\centering
\includegraphics[width=\textwidth]{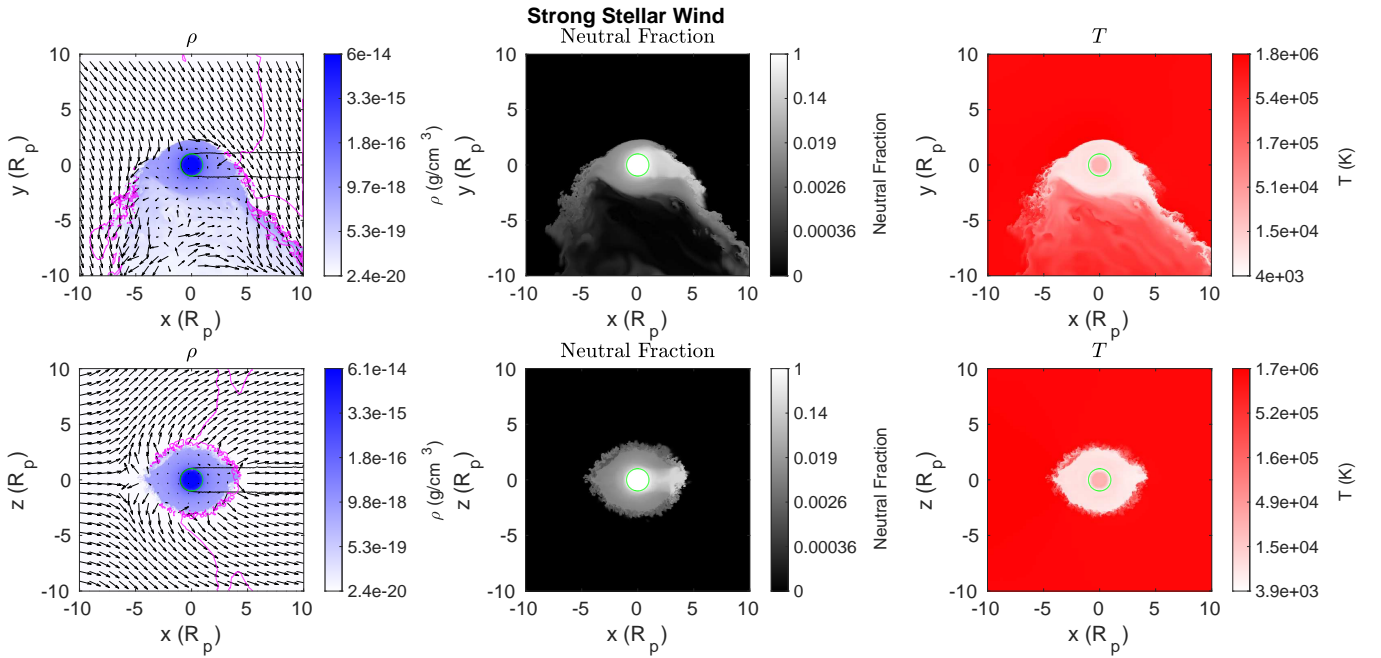}
\caption{Same as figure \ref{fig:rxt_low_wind}, for the high stellar wind case. Here the stellar wind has significantly disrupted both the up-orbit and down-orbit arms, with material leaving the planetary Hill sphere primarily through the L1 and L2 points.}
\label{fig:rxt_high_wind}
\end{figure*}

Figure \ref{fig:rxt_high_wind} shows the steady state of the high-wind run, which uses a stellar wind calculated to significantly affect the initial steady state of the planetary wind. The stellar wind here is strong enough to confine the planetary wind to the vicinity of the planet, as the stellar wind penetrates the planet's Roche lobe, with planetary wind escaping primarily through the L1 and L2 points.

The $\tau = 1$ radius remains unaffected by the introduction of the stellar wind, as should be expected, since the stellar wind is completely ionized. As in the low-wind case, Kelvin-Helmholtz instabilities can be seen along the edges of the wind, particularly in the down-orbit arm. In contrast, the up-orbit arm is fragmented as it is blown down and away from the planet. A much more significant low-density interaction region is seen in the down-orbit direction as a result of material being ablated from the planetary wind. The central panels show that the neutral tail of the planetary wind is interacting strongly with the stellar wind approximately $5 R_p$ directly anti-stellar.

\begin{figure*}
\centering
\includegraphics[width=\textwidth]{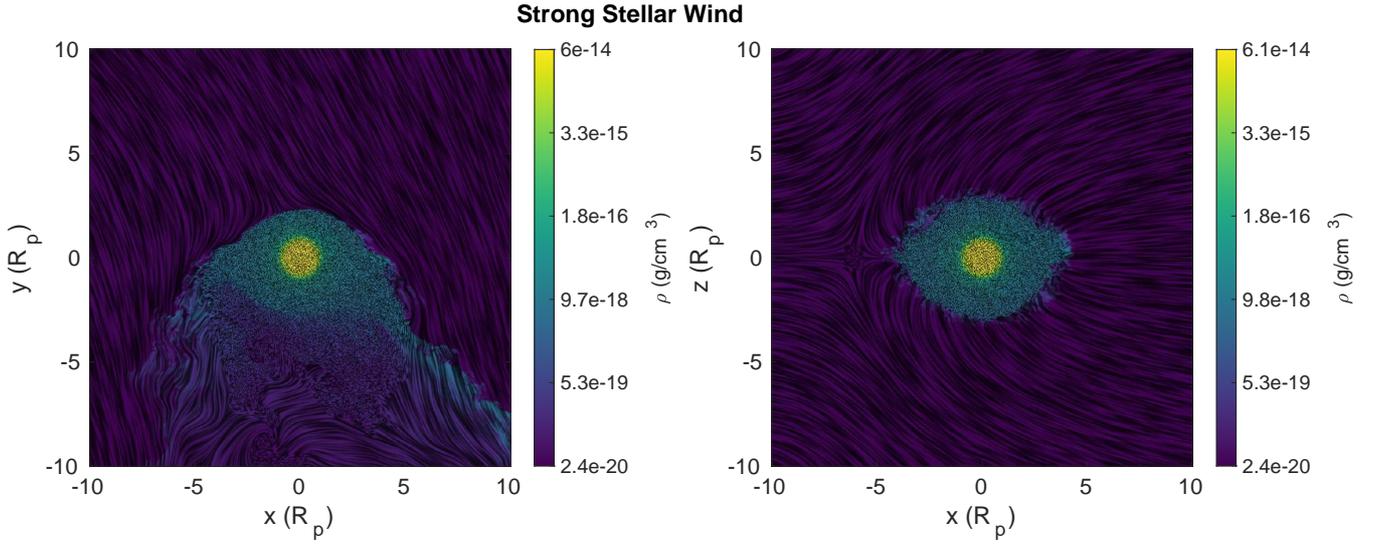}
\caption{Same as figure \ref{fig:flow_texture_low_wind}, for the high stellar wind case.}
\label{fig:flow_texture_high_wind}
\end{figure*}

Figure \ref{fig:flow_texture_high_wind} highlights the confinement of the planetary wind by the stellar wind. The stellar wind flows smoothly around the planet, except for the large interaction region down-orbit of the planet.

\begin{figure*}
\centering
\includegraphics[width=\textwidth]{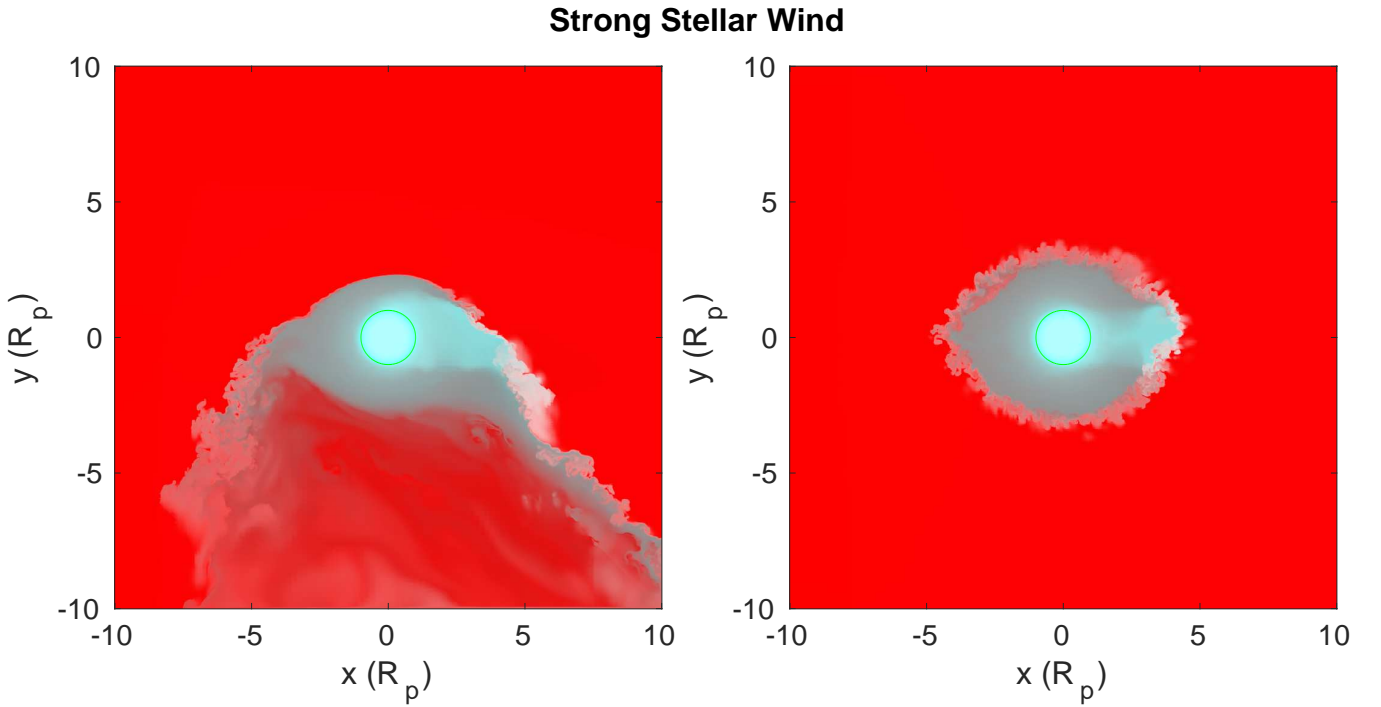}
\caption{Same as figure \ref{fig:charge_exchange_low_wind}, for the high stellar wind case. We see a much larger proportion of white stellar neutral material on the upper edge of the down-orbit arm, where the stellar wind penetrates deeply enough to interact with the neutral tail, as well as more significant numbers of stellar neutrals around the edges of the wind.}
\label{fig:charge_exchange_high_wind}
\end{figure*}

Figure \ref{fig:charge_exchange_high_wind} shows a much larger proportion of high-temperature neutral material, particularly along the up-orbit side of the planetary wind's neutral tail, where the tail and the stellar wind interact. A smaller amount of hot neutral material can also be seen in the remnants of the disrupted up-orbit arm.

\subsubsection{Solar-type stellar wind}

\begin{figure*}
\centering
\includegraphics[width=\textwidth]{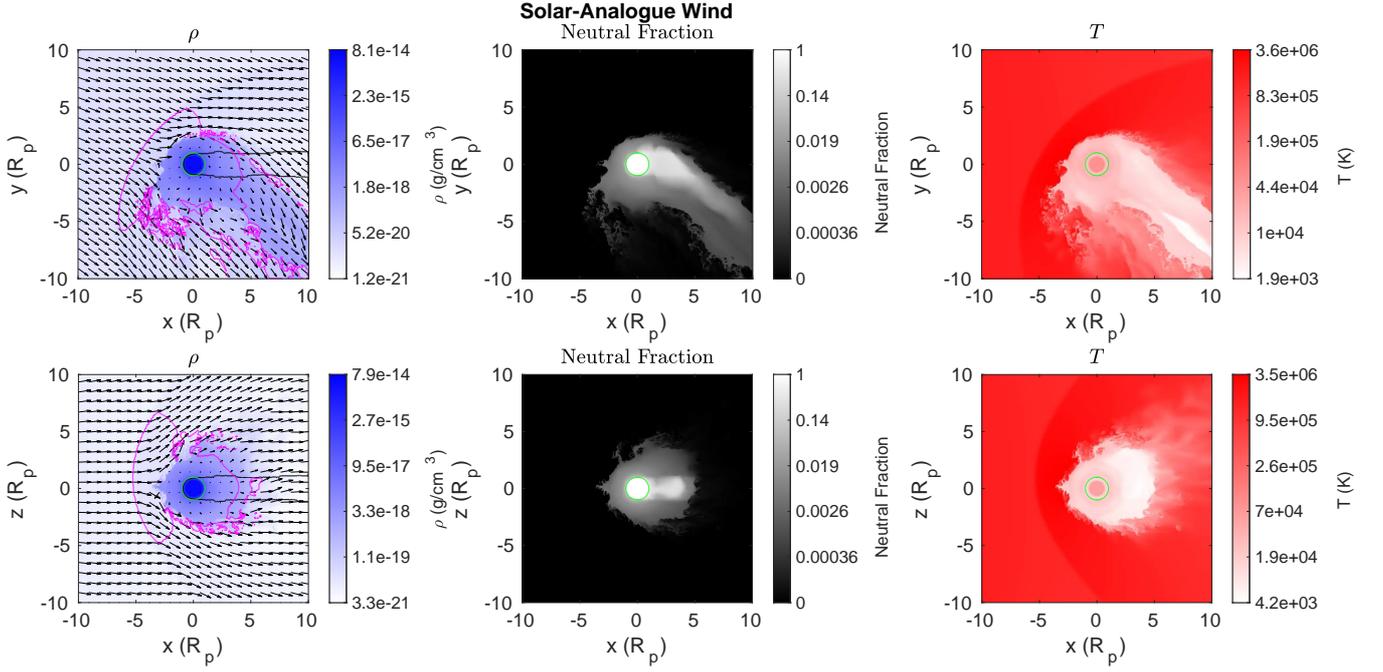}
\caption{Same as figure \ref{fig:rxt_low_wind}, for the solar-analogue stellar wind case. Note the strong bow shock and cometary tail.}
\label{fig:rxt_solar_wind}
\end{figure*}

\begin{figure*}
\centering
\includegraphics[width=\textwidth]{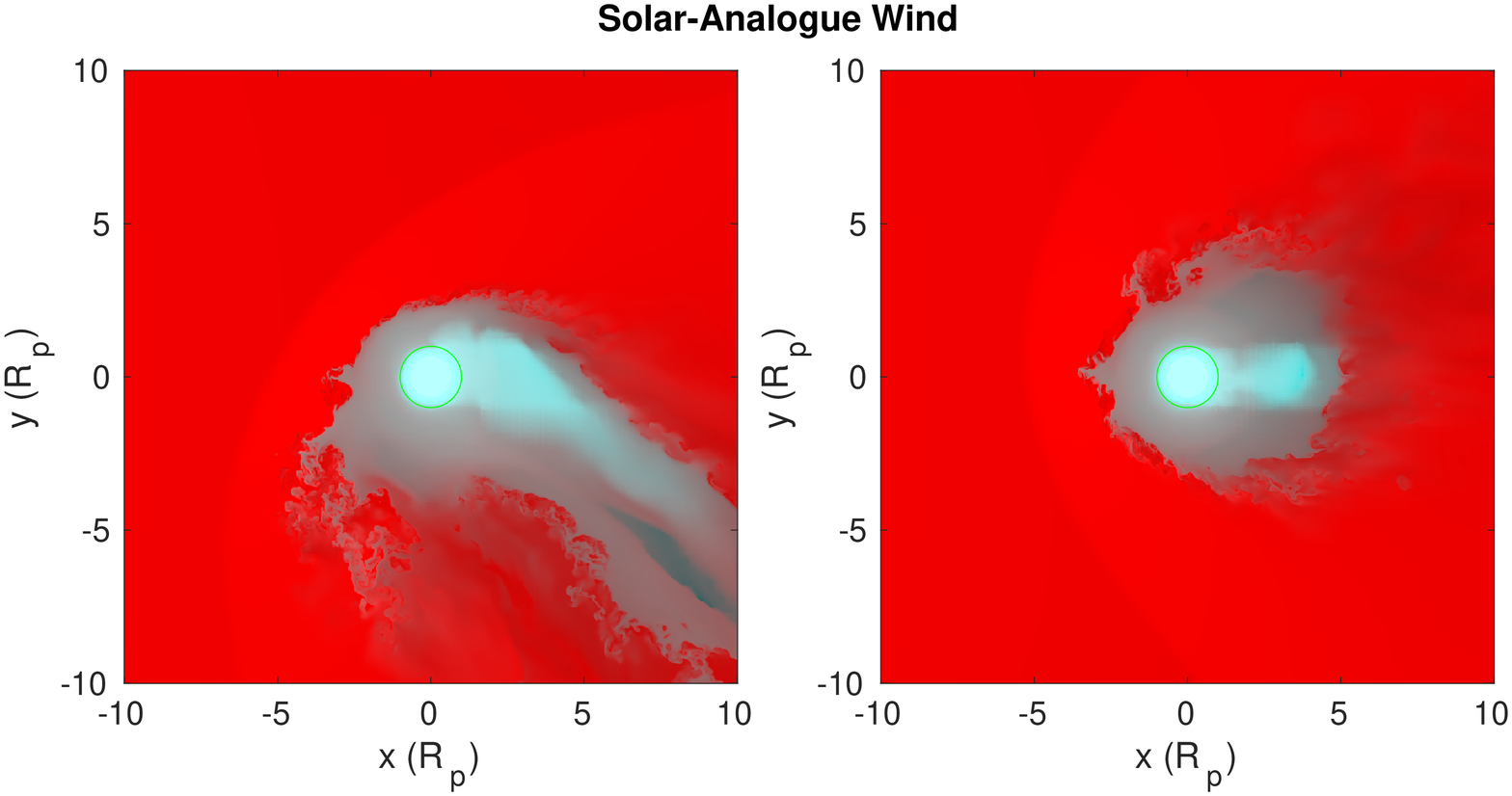}
\caption{Same as figure \ref{fig:charge_exchange_low_wind}, for the solar-analogue stellar wind case. Much of the neutral material is confined to the ionization shadow of the planet.}
\label{fig:charge_exchange_solar_wind}
\end{figure*}

Figure \ref{fig:rxt_solar_wind} shows the final state of the solar-analogue stellar wind. A strong bow shock is apparent where the stellar wind meets the planetary wind. In addition, the ionization front ($\tau = 1$ surface) has been pushed slightly inward from $R_p$. The wind does create a cometary tail, but as figure \ref{fig:charge_exchange_solar_wind} shows, the neutral fraction of the wind is not significantly higher than in the low-wind case; therefore, despite the greater bulk velocity of the neutral material, the high-velocity Lyman-$\alpha$ absorption is not significantly affected.

\begin{figure*}
\centering
\includegraphics[width=\textwidth]{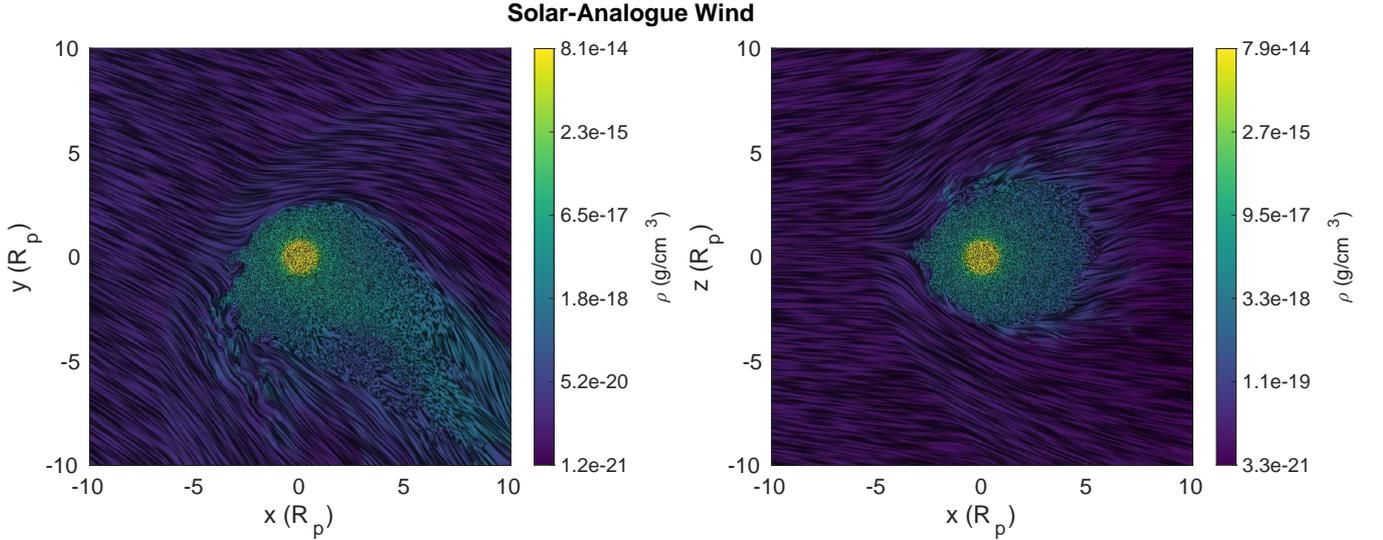}
\caption{Same as figure \ref{fig:flow_texture_low_wind}, for the solar-analogue stellar wind case.}
\label{fig:flow_texture_solar_wind}
\end{figure*}

Figure \ref{fig:flow_texture_solar_wind} illustrates the significantly more radial nature of the stellar wind velocity, as well as the limited region of wind-wind interaction.

\subsection{Charge exchange potential}

\begin{figure*}
\centering
\includegraphics[width=\textwidth]{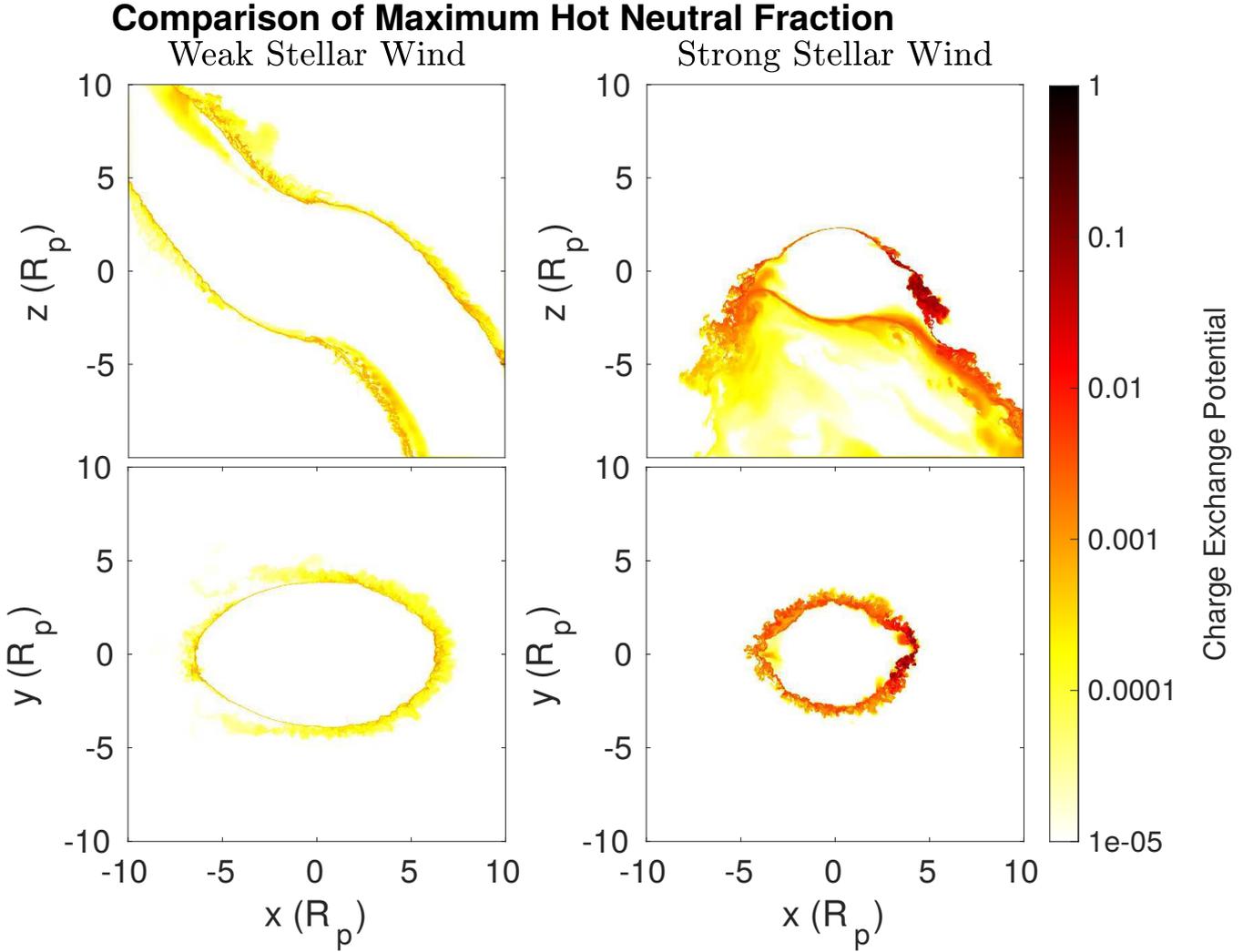}
\caption{The charge exchange potential here is the percentage of the total hydrogen that could theoretically be converted to hot neutral hydrogen. In the weak stellar wind case, the low stellar wind density means that only a very small fraction of the planetary wind can exchange electrons with the stellar wind. On the other hand, in the strong stellar wind case almost 100\% of the hydrogen could be converted to stellar neutrals in some areas where the stellar wind interacts with the neutral tail, and a few percent in other regions.}
\label{fig:charge_exchange_pot}
\end{figure*}

Figure \ref{fig:charge_exchange_pot} is intended to highlight the difference between the potential amount of charge exchange in the low-wind and high-wind cases. The top row shows the top view, the bottom row shows the side view; the left column shows the low-wind case, the right column shows the high-wind case. The charge exchange potential is calculated by determining the fraction of the total material that could be converted to hot neutral material, if there were no reverse reactions (hot neutral interacting with cold ion) taking place: $\min({\nhii}_*, {\nhi}_p)/\nH$.

In the low-wind case, only about 0.01\% of the total hydrogen could be converted in most of the interaction region, with a maximum of about 1\% in a thin layer near the planetary wind. In the high-wind case, in contrast, almost the maximum of 50\% can be converted to hot neutrals in the region where the stellar wind interacts strongly with the neutral tail of the planetary wind, and there are large regions where 0.1-1\% of the hydrogen can become hot neutral material.

This figure also highlights the importance of resolution in the calculation of charge exchange. Turbulence greatly increases the rate of mixing of stellar protons and planetary hydrogen, and a resolution sufficient to resolve the turbulent mixing regions is required to fully describe charge exchange. In the low-wind case, the mixing layer at the edge of the planetary wind is resolved by 15 cells. In the high-wind case, the smallest mixing layer is at the shock between the stellar and planetary winds at the most direct interface, which is resolved by only a couple of cells; outside of this shock, though, the mixing region is resolved by no less than 20 cells. In contrast, \citet{khodachenko19}, for example, resolved their simulations in similar locations only to approximately $0.5 R_p$.

\subsection{Synthetic observations}

We compute synthetic observations for the no-wind, low-wind, and high-wind cases, as described in \citet{carroll16} and \citet{debrecht20}, with the modification that planetary material is held at $10^4$ K and the stellar material is held at $10^6$ K. This is justified by the much larger charge exchange cross section compared to other interactions (see section \ref{sec:cross_sec}).

\begin{figure*}
\centering
\includegraphics[width=\textwidth]{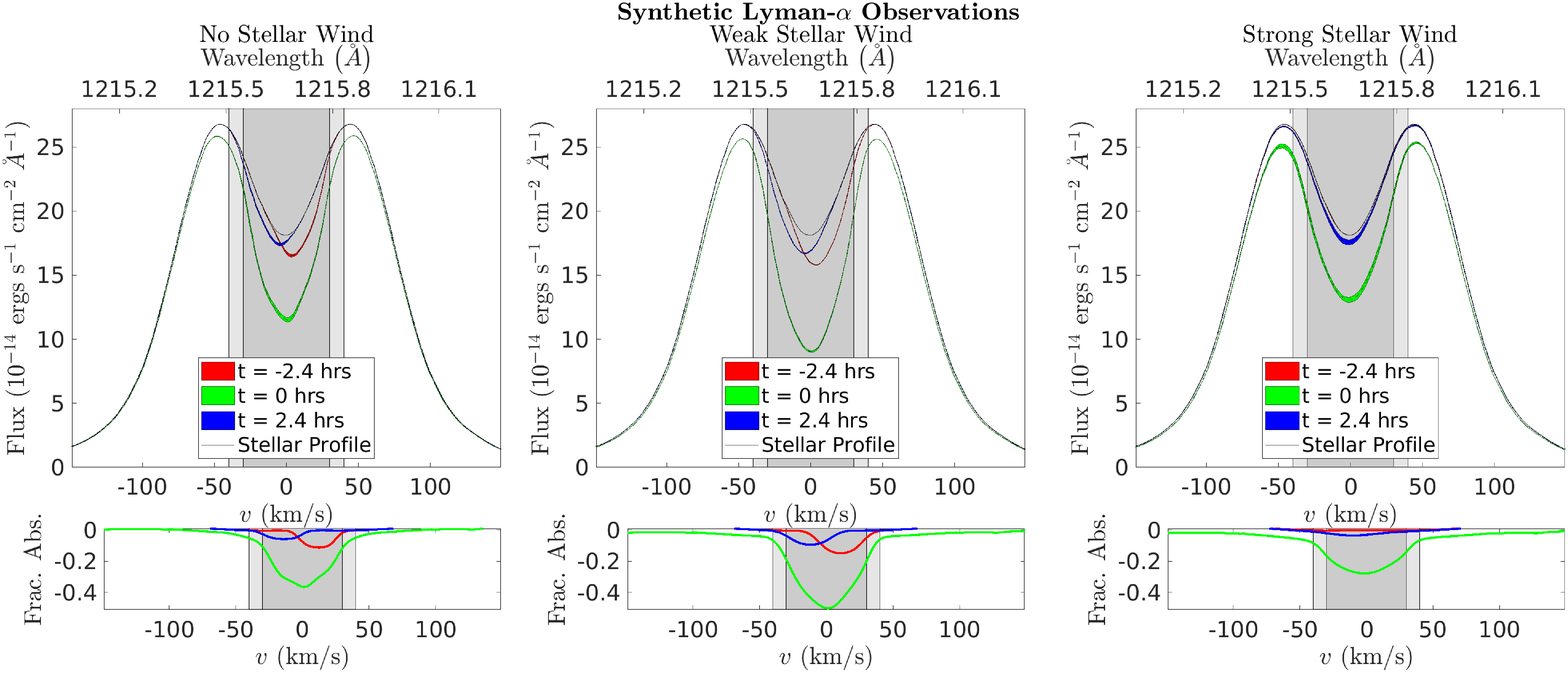}
\caption{Comparison of synthetic observations of the no-wind, low-wind, and high-wind cases. The dark grey and light grey regions represent $\pm30 \mbox{ km/s}$ and $\pm40 \mbox{ km/s}$, respectively, where interstellar absorption and geocoronal emission prevent the detection of planetary absorption signals. Thick lines represent observations within one standard deviation of the mean time-variable absorption, while the dashed line gives the observed average flux. The fractional absorption for each simulation is given in the bottom panels. While the absorption is greatest around line center, there is about 1\% absorption at high velocities in the low-wind case and 2\% absorption at high velocities in the high-wind case. Note that the out-of-transit observations have artificially-lowered obscuration fractions due to the size of our simulation box.}
\label{fig:obs}
\end{figure*}

\begin{figure*}
\centering
\includegraphics[width=\textwidth]{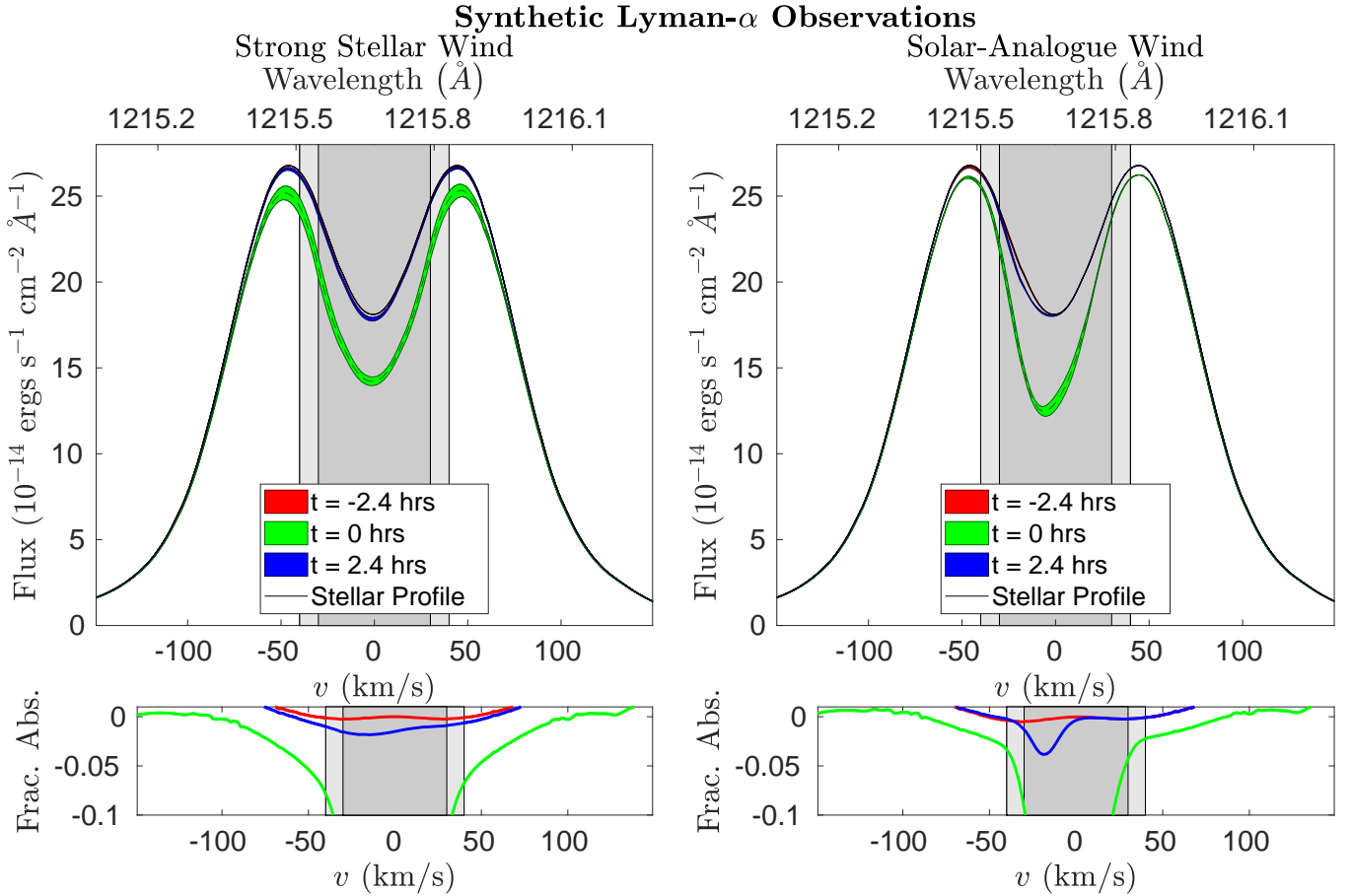}
\caption{Comparison of synthetic observations of the high-wind and solar-analogue-wind cases, as in figure \ref{fig:obs}. While the observable absorption is lower in the solar-analogue case, the ISM-obscured absorption is skewed more strongly toward the blue wing, as expected for a cometary tail. Note that the left-hand panel duplicates the right-hand panel of Figure \ref{fig:obs}.}
\label{fig:obs_solar}
\end{figure*}

In the top panels of figures \ref{fig:obs} and \ref{fig:obs_solar} we show the synthetic Lyman-$\alpha$ observations for these three cases, including their variability over time. As in \citet{debrecht20}, thick lines represent observations within one standard deviation of the mean time-variable absorption, while the dashed line gives the observed average flux. Standard deviations were taken over 24.4, 12.2, and 18.3 hours for the no-wind, low-wind, and high-wind cases, respectively. As with radiation pressure in \citet{debrecht20}, we see lesser variability in the low-wind case thanks to smoothing out of the wind-wind interaction region, versus the wind-ambient interaction, and greater variability in the high-wind case.

See \citet{debrecht20}, section 3.3, for a detailed discussion of the no-wind case. The low-wind case has much deeper absorption at line center than the no-wind case, and absorption of about 1\% out to $-150$ km/s in the blue wind, $125$ km/s in the red wing. The high-wind case has a much shallower absorption at line center, thanks in part to the truncation of the up-orbit arm and in part to the increased spread in absorption due to the larger amount of stellar neutral material. The extended absorption is also slightly deeper in the blue wing, with absorption of about 2\% out to $-150$ km/s.

The bottom panels of figure \ref{fig:obs} show the fractional absorption of the Lyman-$\alpha$ line for each simulation. Again, the absorption is confined primarily to the center of the line, though there is small potentially observable absorption outside of the region covered by the interstellar medium and geocoronal emissions.

\begin{figure*}
\includegraphics[width=\textwidth]{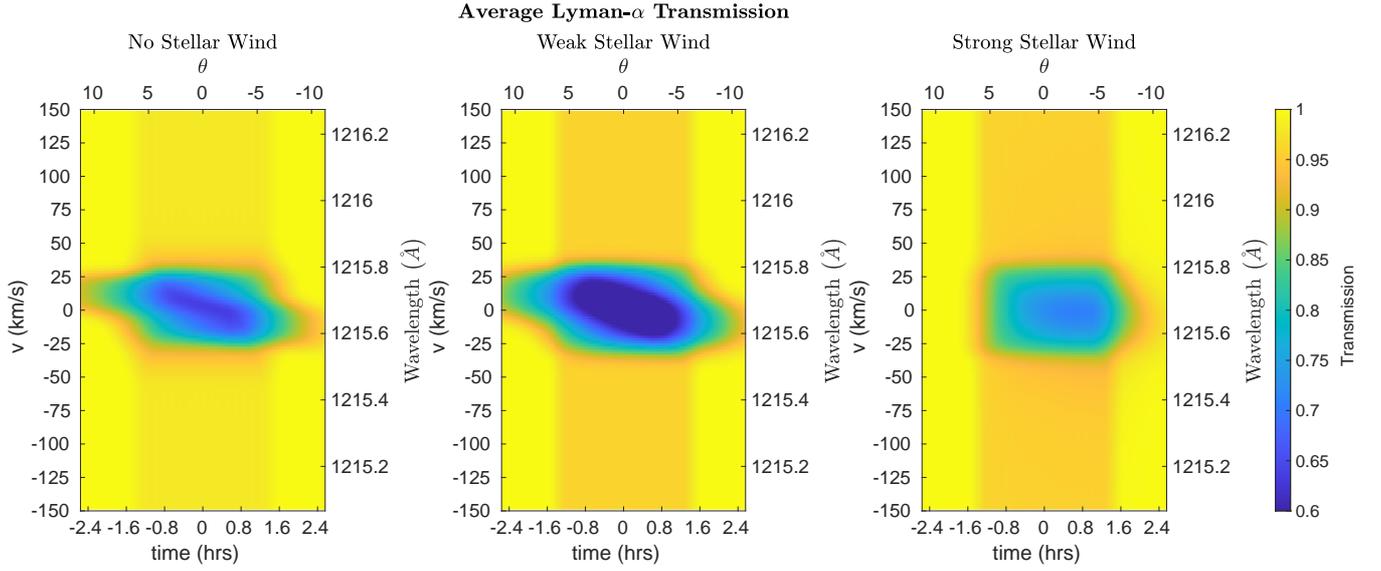}
\caption{Average transmission fraction of the Lyman-$\alpha$ flux for the no-wind, low-wind, and high-wind cases, as a function of time since transit/orbital angle (x axes) and line-of-sight velocity/wavelength (y axes). The truncation of the up-orbit arm causes the transmission fraction to increase sharply 1.5 hours pre-transit. As in figure \ref{fig:obs}, the out-of-transit absorption is artificially lowered due to the size of our simulation box.}
\label{fig:attenuation}
\end{figure*}

Figure \ref{fig:attenuation} is similar to Figure \ref{fig:obs}, but shows the transmission fraction as a function of time after transit (orbital angle) and velocity (wavelength). The truncation of the up-orbit arm in the third panel is particularly apparent here, with a sharp cutoff at 1.5 hours pre-transit. Finally, we note that the Lyman-$\alpha$ absorption found in these simulations is lower than that found in \citet{carroll16}, which is due to a significantly higher ionization fraction for the bulk of the wind in the current simulations, which in turn is due to an ionization timescale in the optically-thin wind of a factor of $\sim 4$ shorter in the current simulations.

\begin{figure*}
\includegraphics[width=\textwidth]{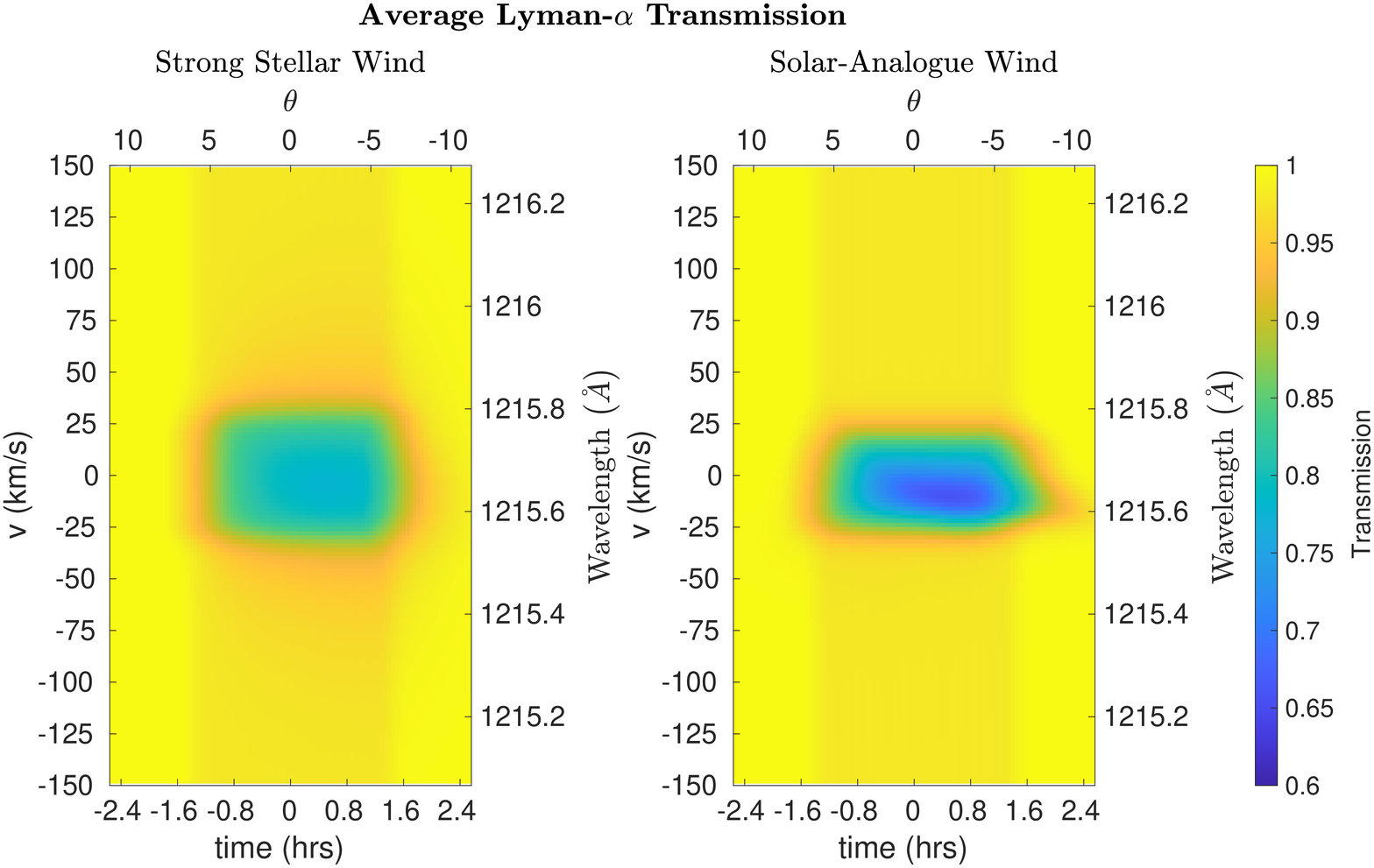}
\caption{Comparison of synthetic observations of the high-wind and solar-analogue-wind cases, as in figure \ref{fig:attenuation}.}
\label{fig:attenuation_solar}
\end{figure*}

Figures \ref{fig:obs_solar} and \ref{fig:attenuation_solar} compare the absorption in the high-wind case and the solar-analogue case. The unobscured absorption is lower in the solar-analogue case.

\section{Discussion}\label{sec:disc}

\subsection{Cross section comparisons} \label{sec:cross_sec}

The cross-sections for the hard-body, proton-proton, and charge exchange interactions differ by approximately an order of magnitude each, with $\sigma_{HI} \approx 10^{-15} \mbox{ cm}^2$, $\sigma_{HII} \approx 10^{-14} \mbox{ cm}^2$, and $\sigma_{CX} \approx 10^{-13} \mbox{ cm}^2$, respectively. Charge exchange therefore happens more rapidly than collisions with ions and neutrals that would thermalize the neutral populations. The mean free path for charge exchange where there are meaningful densities of both hot and cold neutrals is about $0.1 R_p$ (whereas we effectively resolve $\sim0.01 R_p$). The cross section for neutral-neutral interactions is about $10 R_p$. Therefore, there is the potential for charge exchange to occur well before the stellar and planetary populations thermalize. However, comparing the synthetic observations with the planetary and stellar populations held at constant temperatures of $10^4$ and $10^6$, respectively, to observations where the expected temperature at local equilibrium is used, we find no discernible difference.

\subsection{Comparison with the analytic treatment}\label{sec:cptoan}

Our approximate analysis (section \ref{sec:math}) showed that we should expect a maximum number of approximately $40 \mbox{ cm}^{-3}$ stellar neutrals for the weak stellar wind and $2\times10^3 \mbox{ cm}^{-3}$ stellar neutrals for the strong stellar wind. We actually find maxima of $\sim 1 \mbox{ cm}^{-3}$ and $\sim 10^4 \mbox{ cm}^{-3}$, which agree reasonably well. Note that in the weak stellar wind case, the stellar wind doesn't interact significantly with the neutral tail, which explains why the maximum is lower than expected. On the other hand, the stellar wind tends to compress the neutral tail, leading to greater density for the stellar wind to interact with, which explains why the maximum is higher than expected.

\subsection{Comparison to Previous Work} \label{sec:comp}

For our high stellar wind simulation, we use the properties of the intermediate wind from \citet{mccann18}. Their study shows that a more highly inflated planet, when encountering a similar wind, has an intermittent "burping" phenomenon, in contrast to our nearly confined wind, primarily due to the increased pressure of the planetary wind. We also note the similarity to the simulations of \citet{cherenkov18}, whose simulations also showed planetary escape that was confined to near the L1 and L2 points. Here the up-orbit arm is more strongly truncated than in their simulations, thanks to a greater stellar wind pressure.

As in most of the simulations of HD~209458b with charge exchange, including those of \citet{khodachenko17}, \citet{shaikhislamov16}, \citet{christie16}, and \citet{esquivel19} (in the absence of radiation pressure), we find that the hot neutral material (a.k.a. ENAs) absorbs only an extra few percent of the high-velocity Lyman-$\alpha$ line. Using a much higher stellar mass loss rate, \citet{tremblin13} found an absorption closer to that seen in observations. This suggests that a greater density at the same radius, wind temperature, and wind velocity, with the resulting greater potential for charge exchange, could lead to the levels of absorption expected from observations of HD~209458b.

This also suggests one reason why models of GJ~436b have in general been successful in reproducing the observed absorption features, while simulations of HD~209458b have not. HD~209458b's escape velocity at the UV absorption radius is 41.6 km/s, while GJ~436b's escape velocity is only slightly more than half that, at 24.6 km/s. At similar levels of EUV flux, GJ~436b will therefore have a higher mass loss rate, if we assume equal efficiency in converting deposited stellar energy to mass loss. This in turn allows for greater production of ENAs, assuming similar stellar wind environments.

\section{Conclusion} \label{sec:conc}

Based on our synthetic observations, we find that the stellar winds investigated here produce insufficient hot neutral hydrogen (ENAs) when interacting with the planetary wind of HD~209458b to produce the expected high-velocity absorption features. Previous studies have shown that it may be possible to retrieve the expected absorption from HD~209458b with higher stellar mass loss rates \citep{tremblin13}, or by combining the effects of charge exchange and radiation pressure \citep{esquivel19}. In addition, combinations of stellar and planetary winds from GJ~436b have been found that reproduce similar high-velocity absorption features, some without the inclusion of charge exchange. The investigation of the interaction of GJ~436b with its host star's wind is left for future work.

\section{Acknowledgments}

We thank the Other Worlds Laboratory (OWL) at University of California, Santa Cruz for facilitating this collaboration by way of the OWL Exoplanets Summer Program, funded by the Heising-Simons Foundation. This work used the computational and visualization resources in the Center for Integrated Research Computing (CIRC) at the University of Rochester and the computational resources of the Texas Advanced Computing Center (TACC) at The University of Texas at Austin, provided through allocation TG-AST120060 from the Extreme Science and Engineering Discovery Environment (XSEDE) \citep{xsede}, which is supported by National Science Foundation grant number ACI-1548562. Financial support for this project was provided by Department of Energy grants DE-SC0001063, DE-SC0020432, and DE-SC0020434, the National Science Foundation grants AST-1515648, AST-1813298, and AST-1411536, and the Space Telescope Science Institute grants HST-AR-12832.01-A and HST-AR-14563.001-A. EB acknowledges additional support from KITP UC Santa Barbara, funded by NSF Grant PHY-1748958, and the Aspen Center for Physics, funded by NSF Grant PHY-1607611.

\section{Data Availability}

The data underlying this article will be shared on reasonable request to the corresponding author.

\clearpage
\balance

\bibliography{planets.bib}

\bsp

\label{lastpage}

\end{document}